\documentclass[runningheads]{llncs}
\usepackage{graphicx}
\usepackage[utf8]{inputenc}
\usepackage{enumitem}
\usepackage{xspace}
\usepackage{mathtools}
\usepackage{nicefrac}
\usepackage{stmaryrd}
\usepackage{amsmath}
\usepackage{amssymb}
\usepackage{nccmath}
\usepackage{algorithm}
\usepackage{algorithmic}

\DeclareMathOperator*{\argmax}{arg\,max}
\DeclareMathOperator*{\argmin}{arg\,min}

\newcommand{\size}[1]{\ensuremath{\llbracket #1 \rrbracket}}

\newtheorem{informalclaim}{Informal Claim}

\usepackage{pgf, tikz}
\usetikzlibrary{arrows, automata}
\usetikzlibrary{arrows,fit}
\usetikzlibrary{positioning}
\usetikzlibrary{calc}
\usetikzlibrary{shapes,decorations}
	\tikzset{
		inner sep=0pt, outer sep=0pt, minimum size=0pt, thick,
		level/.style={sibling distance = (\columnwidth/16)*2^(4-#1)},
		winner/.style={minimum size=1.5em, circle, draw, fill=white, font={\footnotesize}},
		split/.style={minimum size=1.5em, inner sep=1pt, circle split, draw, fill=white, font={\tiny}},
		leaf/.style={inner sep=.15em, font={\footnotesize}},
		ball/.style={minimum size=.4em,circle,fill=black},
		beats/.style={thick,->,>=stealth',draw},
		tline/.style={thick,draw}
	}

\begin{document}

\title{In the Beginning there were $n$ Agents:\\Founding and Amending a Constitution}
\titlerunning{In the Beginning There Were $n$ Agents}

\author{Ben Abramowitz\inst{1,2} \and Ehud Shapiro\inst{2} \and Nimrod Talmon\inst{3}}
\institute{Rensselaer Polytechnic Institute, Troy NY 12180, USA \email{abramb@rpi.edu} \and Weizmann Institute of Science, Rehovot, Israel \email{ehud.shapiro@weizmann.ac.il} \and Ben-Gurion University, Be'er Sheva, Israel \email{talmonn@bgu.ac.il}}

%
%

\maketitle

\begin{abstract}
Consider $n$ agents forming an egalitarian, self-governed community. Their first task is to decide on a decision rule to make further decisions.
We start from a rather general initial agreement on the decision-making process based upon a set of intuitive and self-evident axioms, as well as simplifying assumptions about the preferences of the agents. From these humble beginnings we derive a decision rule. Crucially, the decision rule also specifies how it can be changed, or amended, and thus acts as a \emph{de facto} constitution.
Our main contribution is in providing an example of an initial agreement that is simple and intuitive, and a constitution that logically follows from it.
The naive agreement is on the basic process of decision making --  that agents approve or disapprove proposals;  that their vote determines either the acceptance or rejection of each proposal; and on the axioms, which are  requirements regarding a constitution that engenders a self-updating decision making process.
\end{abstract}


\section{Introduction}

Consider a group of $n$ agents that gather to form a self-governed community. To make collective decisions, they first need to agree on a rule by which to do so; but offering to vote on the voting rule leads to infinite regress.

Thus, in this paper we offer a different approach.
In particular, 
we assume that the agents first agree on a common set of axioms, and aim to infer an initial decision rule from these axioms.  Ideally, the axioms will logically imply an identifiable set of permissible decision rules, and a unique decision rule from this set to serve as the initial rule, so that an agent who agrees to the axioms cannot object to the initial rule they imply, nor to the set of permissible rules.

Furthermore, it is crucial that the decision rule be amendable. Once the agents have passed the first hurdle and founded their rudimentary constitution, agent preferences over the various permissible decision rules become material. 
Thus, we assume that the agents also agree on a common set of axioms that relate to the amendment procedure to change the decision rule itself. Again, ideally, the axioms will logically imply a unique amendment procedure such that, taking agent preferences over possible decision rules into account, agents who agree to the axioms would have no objections to the amendment procedure.

We view the initial decision rule as the foundation of the \emph{constitution} of the community and the amendment procedure as the process by which the constitution changes itself. 
We look at our work as a step towards understanding how a group of agents can congeal into a self-governed, rule-based, egalitarian community with a decision making mechanism that is not dictated by an external agent.

\subsection{Paper structure}

We discuss some related work in Section~\ref{section:related work}.
In Section~\ref{section:informal} we discuss how to found and amend a constitution; we do so in an informal way, to mimic the way by which the agents may freely discuss the principles by which they wish to be self-governed, and to introduce the intuitions behind the formal treatment that follows.
In Section~\ref{section:formal} we formally define the set of axioms and their logical consequences -- a unique decision rule and amendment procedure.
In Section~\ref{section:alternative} we present an alternative axiomatic basis, explore its consequences, and compare it to the initial axiomatic basis proposed. 
We conclude, in Section~\ref{section:discussion}, with a discussion on future research directions.

\section{Related Work}\label{section:related work}
Kenneth May's seminal paper~\cite{may1952set} demonstrates that majority rule is the only voting rule that satisfies a compact, intuitive set of axioms. We have taken May's Theorem as inspiration to answer the question of how a group of agents could establish a set of rules for themselves where none existed before. While our axioms lead the agents to use majority rule initially when founding their constitution, they also enable the agents to change this rule according to their collective preferences.
While there is a large literature on characterizing social choice rules axiomatically (see, e.g.,~\cite{campbell2002impossibility,sen2017collective}) and implementation theory (see, e.g.,~\cite{jackson2001crash}), and a small body of work on voting on criteria~\cite{nurmi2015choice,de2019criterion,suzuki2020characterization}, we believe that our work is the first to introduce a set of axioms that simultaneously lead to the formulation of an initial rule and the process by which it can amend itself.
Perhaps the closest work to this comes from the study of legislative procedures and procedural choice~\cite{diermeier2020self,diermeier2015procedural}
We consider two alternative axioms for how the agents can determine what proposal should win if multiple incompatible proposals are viable, namely Condorcet-consistency and Conservatism. Extensions of the Condorcet principle have a long history in social choice \cite{fishburn1977condorcet,brandt2018extending}, and the principle of Conservatism comes from reality-aware parameter update rules \cite{shahaf2018sybil}. Our axiom of Minimalism and assumptions about agent preference structure overcome the general impossibility result of~\cite{houy2004note} without concern of infinite regress~\cite{suzuki2015order}.

One line of research on constitutional amendments assumes that voter preferences over possible rules are derived from their preferences over the outcomes that each rule produces~\cite{koray2000self}. This is the ``consequentialist approach'' used in game theoretic  analysis~\cite{lagunoff2009dynamic}. Perhaps the work here most similar to ours is that of Bhattacharya~\cite{bhattacharya2019constitutionally} as they consider single-peaked preferences.
We take a different approach to modeling preferences over constitutional amendments, more similar to that of Barbera and Jackson~\cite{barbera2004choosing}. In their model, constitutions are just a set of voting rules. A constitution may be a single rule used to update itself, or it can consist of a fixed amendment rule used to update the decision rule.
In contrast to Barbera and Jackson, who conclude that a rule used to update itself is too simple for their setting, we see this simplicity as a virtue in bootstrapping a constitution.

\section{Founding and Amending a Constitution, Informally}\label{section:informal}

We begin by informally discussing the processes of founding and amending a constitution; later, we delve into a rigorous mathematical treatment.

\subsection{Founding a Constitution, Informally}\label{section:founding informal}

We assume that all $n$ agents agree that they need a procedure for accepting or rejecting proposals.
Proposals are considered individually, and for each proposal, each agent either approves it or not. 
The agents wish to establish a decision rule that determines whether the group as a whole accepts or rejects a given proposal, based on the votes of its members.  To achieve that, they first agree on the following axioms that a decision rule must fulfill.

We first introduce the axioms without mathematical precision, to demonstrate how unassuming they are, and show that the claims that follow from them make intuitive sense. In the next Section~\ref{section:formal} we rephrase the axioms mathematically, rewrite the claims as propositions, and prove the propositions formally.

\subsubsection*{Axioms for a Decision Rule}
\begin{enumerate}[itemsep=1ex]

\item \label{axiom:decisiveness}
\textbf{Decisiveness}: For a given proposal, the decision is either to accept or to reject it, as determined by the votes.

\item \label{axiom:monotonicity}
\textbf{Monotonicity}: A decision to accept a proposal will remain so if some agents that do not approve the proposal change their vote to approve.

\item \label{axiom:anonymity}
\textbf{Anonymity}: A decision to accept a proposal shall not depend on the identities of the agents approving it.

\item \label{axiom:concordance}
\textbf{Concordance}: If two proposals are incompatible, in the sense that no agent may approve both, then not both can be accepted.

\item \label{axiom:bootstrap}
\textbf{Minimality}: Among the decision rules consistent with the axioms above, choose an initial rule that requires a minimal number of approvals.
\end{enumerate}

Axioms~\ref{axiom:decisiveness}-\ref{axiom:anonymity} imply that the decision rule is based on the size of the fraction of the agents approving the proposal.

\begin{informalclaim}[Fractional Approval Voting]\label{claim:first} 
Axioms~\ref{axiom:decisiveness}-\ref{axiom:anonymity} imply that the decision rule shall accept a proposal if approved by at least some fraction of the agents. 
\end{informalclaim}

Adding Axiom~\ref{axiom:concordance} implies that the fraction shall be at least a half.

\begin{informalclaim}[Supermajority Approval]
Axioms~\ref{axiom:decisiveness}-\ref{axiom:concordance} imply that the fraction of approvals needed to accept a proposal is greater than $\frac{1}{2}$. 
\end{informalclaim}

We refer to a fraction of approvals greater than $\delta \ge \frac{1}{2}$ as a \emph{$\delta$-supermajority}, with the case of $\delta = \frac{1}{2}$ being \emph{simple majority} and $\delta = 1 - \epsilon$, where $0 < \epsilon  \leq \frac{1}{n}$, being \emph{unanimous consent}. 

Note that, unlike Axioms~\ref{axiom:decisiveness}-\ref{axiom:concordance}, which effectively restrict the set of possible decision rules, Axiom~\ref{axiom:bootstrap} compares and prioritizes some decision rules over others. 

\begin{informalclaim}[Initially, Simple Majority Approval]\label{claim:last}
Axioms~\ref{axiom:decisiveness}-\ref{axiom:bootstrap}  imply that the initial decision rule is approval via a simple majority.
\end{informalclaim}

To summarize, we claim that, according to Axioms~\ref{axiom:decisiveness}-\ref{axiom:bootstrap}, the decision rule must always be a $\delta$-supermajority approval rule, $\frac{1}{2} \le \delta < 1$, with the initial supermajority being a simple majority $\delta = \frac{1}{2}$.

\subsection{Amending a Constitution, Informally}\label{section:amending informal}

The initial decision rule of the community is simple majority according to Axiom~\ref{axiom:bootstrap}, but the axioms are consistent with amending it to any $\delta$-supermajority, with $\frac{1}{2} \le \delta < 1$. So, once the initial rule is established as simple majority, some agents may wish to amend it to a higher supermajority rule. Some agents may aspire for a decision by unanimous consent, while others may prefer to require more than simple majority but less than unanimous consent, for example to protect minorities. Our assumption that agents may prefer greater supermajorities for decisions is consistent with analytical and empirical research that suggests this often holds even for agents in the majority~\cite{diermeier2015procedural}. Agents may also wish to employ different rules for different decisions in the future.
Furthermore, agents' preferences may change over time. Thus, below we focus on the fundamental question of how the decision rule can be amended.

Naturally, the axioms agreed upon by the community apply to the decision of whether to accept a newly-proposed decision rule just like they would for any other decision.
Hence, the decision to change the present $\delta$-supermajority decision rule must be approved by a $\delta$-supermajority.
However, proposals to amend the decision rule itself require additional considerations, reflected in the consistency requirements introduced below. But first we provide some background in order to express them.

We assume that agents have preferences over decision rules, and that an agent approves a proposal to amend the decision rule if and only if the agent prefers the newly-proposed rule over the decision rule in force.  
Concretely, we identify agents with their index $i \in [n]$, and make the simplifying assumption that agent preferences are single-peaked in the following sense: Every agent $i \in [n]$ has an \emph{ideal point} $\delta_i$, $\frac{1}{2} \le \delta_i < 1$, such that: 
(i) agent $i$ strictly prefers $\delta_i$ over all other proposed $\delta$'s;
(ii) among any two proposed values larger than $\delta_i$, the agent prefers the smaller of the two;
(iii) among any two proposed values smaller than $\delta_i$, the agent prefers the larger of the two; 
and (iv) the agent $i$ has no other pairwise preferences.

Secondly, we say that a proposal $p$ is \emph{preferred} over another proposal  (aka \emph{dominates}) $p'$ if the set of agents that prefer $p$ over $p'$ are a majority.
A proposal $p$ is \emph{most-preferred} if no other proposal $p'$ is preferred over $p$.
Note that, if $p$ is the only most-preferred proposal, then it is a Condorcet winner \cite{brandt2016handbook}. Also, as preference is transitive and there are at most $\frac{n}{2}$ significant values of $\delta$  ($\frac{1}{2},  \frac{1}{2}+\frac{1}{n}, \ldots, \frac{n-1}{n}$), a knock-out tournament among these values for $\delta$, choosing the preferred value in each match, may quickly result in a most-preferred value for $\delta$.

Using these terms, we express two amendment axioms.  The first axiom, named
Posterior Consistency, says that,  to change the current decision rule to a new decision rule,  this amendment decision must be justifiable in retrospect according to the new rule, as opposed to being consistent only with current rule in force. This is in essence an ``anti-hypocrisy'' axiom. The second axiom requires that among all alternative decision rules consistent with the other axioms, we choose a most-preferred decision rule.

\subsubsection*{Axioms for a Decision Rule Applied to Amend Itself}
\begin{enumerate}[itemsep=1ex]
\setcounter{enumi}{5}
\item \label{axiom:posterior}
\textbf{Posterior Consistency}:  A proposal to amend the decision rule is accepted only if accepted according to the newly accepted decision rule as well.

\item \label{axiom:condorcet}
\textbf{Condorcet Consistency}: Among the decision rules that satisfy all axioms above, choose a most-preferred one.
\end{enumerate}


Given our assumptions on agent preferences and Axioms~\ref{axiom:decisiveness}-\ref{axiom:bootstrap}, we claim that there is a unique amendment process that is also consistent with Axioms 1-\ref{axiom:condorcet}

\begin{informalclaim}[Condorcet Amendment Rule]\label{claim:condorcet}
Axioms~\ref{axiom:decisiveness}-\ref{axiom:condorcet} together with our assumptions on agent preferences imply that the process to amend the current $\delta$-supermajority approval decision rule must be:
\begin{enumerate}[itemsep=1ex]

\item Increase $\delta$ to $\delta'>\delta$, if and only if there is a $\delta'$-supermajority that prefers $\delta'$ over $\delta$ and $\delta'$ is the \textbf{maximal} supermajority with this property; 
\item Decrease $\delta$ to $\delta'<\delta$, if and only if there is a $\delta$-supermajority that prefers $\delta'$ over $\delta$ and $\delta'$ is the \textbf{minimal} supermajority with this property; and 
\item  Retain the present $\delta$-supermajority rule otherwise.
\end{enumerate}
\end{informalclaim}

This completes the informal presentation.

\section{Founding and Amending a Constitution, Formally}\label{section:formal}

In this section we rephrase the assumptions, axioms, and claims formally.

\subsection{Founding a Constitution, Formally}\label{section:founding formal}

Let $B = \{0,1\}$;  these values represent approving or disapproving a given proposal by each agent, as well as accepting or rejecting the proposal by the group of agents as a whole.
For the set of $n$ agents, we refer to $V \in B^n$ as a voter \emph{profile}. For a profile $V$, we define $\size{V} := |\{i ~|~ V_i = 1\}|$ to be the number of approvals in $V$.
For two profiles $V, V' \in B^n$, define $V \le V'$ if $V_i \le V'_i$ for all $1\le i \le n$.

Indeed, we assume a set of $n$ agents that wish to agree on a decision rule $d$ that takes a proposal and a profile $V$, where each voter specifies either \emph{approve} (1) or \emph{not approve} (0); and produces (i.e., outputs) a decision of either to \emph{accept} (1) or to \emph{reject} (0).
In the axioms below, $V, V' \in B^n$ and the given proposal is assumed and not specified as an explicit parameter.

\subsubsection*{Formal Axioms for a Decision Rule}
\begin{enumerate}[itemsep=1ex]
\setcounter{enumi}{8}
\item \label{faxiom:decisiveness}
\textbf{Decisiveness}: The decision rule $d$ is a function $d: B^n \xrightarrow{} B$.
\item \label{faxiom:monotonicity}
\textbf{Monotonicity}: If $d(V) = 1$ and $V \le V'$ then $d(V') = 1$.

\item \label{faxiom:anonymity}
\textbf{Anonymity}: If $V'$ is a permutation of $V$ then $d(V) = d(V')$.

\item \label{faxiom:concordance}
\textbf{Concordance}: If $V_i + V'_i \le 1$ for all $1\le i \le n$, then $d(V) + d(V') \le 1$.

\item \label{faxiom:bootstrap}
\textbf{Minimality}: If $d$ and $d'$ are consistent with the axioms above, and $d'(V') \le d(V)$ implies $\size{V'} \le \size{V}$ for all $V, V' \in B^n$, but not vice versa (exchanging $d$ and $d'$), then prefer $d'$ over $d$.
\end{enumerate}

We now rephrase Informal Claims~\ref{claim:first}-\ref{claim:last} into propositions, and prove them.

\begin{proposition}[Fractional Approval Voting] \label{proposition:fractional}
Axioms~\ref{faxiom:decisiveness}-\ref{faxiom:anonymity} imply that there is some fraction $0 \le f < 1$ such that $d(V) = 1$ if and only if $\size{V}/n > f$.
\end{proposition}

\begin{proof}
Let $V^* \in B^n$ be a voter profile $V^* = \argmin\limits_{\substack{V \in B^n \\ d(V) = 1}} \size{V}$, and let $f = \frac{\size{V^*}-1}{n}$. Naturally, for any profile $\size{V'} < \size{V^*}$, $d(V') = 0$ by the definition of $V^*$. Let $V'$ be a profile for which $\size{V^*} \le \size{V'}$ and hence $f < \size{V^*}/n \le \size{V'}/n$.  We wish to show that $d(V') = 1$.  Consider a voter profile $V'' \in B^n$ such that $V'' \le V'$ and $\size{V''} = \size{V^*}$.  By anonymity, $d(V'') = d(V^*)$ and  by monotonicity, $d(V'') \geq d(V') = 1$, assuming of course that $d$ is a function $d: B^n \rightarrow B$ (Axiom~\ref{faxiom:decisiveness}). 
\qed\end{proof}

With $n$ voters, a decision rule $d$ using fraction $f = \frac{k}{n}$ for $k \in \mathbb{N}$ will return the same outcome as any decision rule $d'$ using fraction $\frac{k}{n} + \epsilon$ for $\epsilon < \frac{1}{n}$, for all possible profiles. We therefore do not differentiate between these functionally equivalent rules. For example, the decision rule using fraction $f = \frac{1}{2}$ is majority rule whether the number of voters is even or odd.

\begin{proposition}[Supermajority Approval]\label{proposition:supermajority}
If $d(V) = 1$ then $\size{V} > \frac{n}{2}$.
\end{proposition}
\begin{proof}
By way of contradiction, assume $d(V) = 1$ and $\size{V} \le \frac{n}{2}$.  Then $V$ has a non-overlapping  permutation $V'$ such that $V_i + V'_i  \le 1$ for all $i \in [n]$. By Axiom~\ref{faxiom:anonymity}, $d(V') = d(V)$, implying via the assumption that $d(V) + d(V') = 2$, contradicting Axiom~\ref{faxiom:concordance}.
\qed\end{proof}

\begin{proposition}[Initially, Simple Majority Approval]\label{proposition:bootstrap}
Axioms~\ref{axiom:decisiveness}-\ref{axiom:bootstrap}  imply that the initial decision rule is $d(V) = 1$ if $\size{V}/n > \frac{1}{2}$.
\end{proposition}
\begin{proof}
Let $d$ be the initial decision rule. By Proposition~\ref{proposition:fractional} it is fractional, with some fraction $0 \le f < 1$.
By Proposition~\ref{proposition:supermajority}, $f \geq \frac{1}{2}$. We claim that $f \leq \frac{1}{2}$.  By way of contradiction, assume $f = \frac{1}{2} + \frac{1}{n}$, and consider a decision rule $d'$ with a fraction $f' = \frac{1}{2}$. Just like $d$, the rule $d'$ is consistent with all the Axioms~\ref{faxiom:decisiveness}-\ref{faxiom:bootstrap}.  Furthermore, it satisfies that $d'(V') \le d(V)$ implies $\size{V'} \le \size{V}$, as $f < f'$.  However, the converse is not true.  Assume $\size{V}=\frac{1}{2}+\frac{1}{n}$ and $\size{V'}=\frac{1}{2}$. Then  $d(V) \le d'(V')$ as both equal 0, 
but $\size{V} > \size{V'}$, contradicting Axiom~\ref{faxiom:bootstrap}.  Hence the initial decision rule is fractional with $f = \frac{1}{2}$, namely approval by a simple majority.
\qed\end{proof}

Let $\Delta := [\frac{1}{2},1)$.  
To summarize, Propositions~\ref{proposition:fractional}-\ref{proposition:bootstrap} prove that, according to Axioms~\ref{faxiom:decisiveness}-\ref{faxiom:bootstrap}, the decision rule must always be a $\delta$-supermajority approval rule, $\delta \in \Delta$, with the initial supermajority being a simple majority $\delta = \frac{1}{2}$.
Next, we rewrite Axioms~\ref{axiom:posterior}-\ref{axiom:condorcet}(a, b) formally, recast Claims~\ref{claim:condorcet}-\ref{claim:equivalence} as proper theorems, and prove them.

\subsection{Amending a Constitution, Formally}\label{section:amending formal}

We have established that all decision rules are $\delta$-supermajority rules for some $\delta \in \Delta$. We therefore identify each decision rule $d$ with its $\delta$, and denote it by $d^\delta$. 


\subsubsection*{Agent preferences.}
Recall that each agent $i$ has a preference $\preceq_i \subseteq  \Delta \times \Delta$ over the decision rules, where $x \prec_i y$ if $x \preceq_i y$ but not vice versa.  
Furthermore, agent preferences are single-peaked in that every agent $i \in [n]$ has an \emph{ideal point} $\delta_i \in \Delta$, such that: 
(i) $\delta \prec_i \delta_i$ for all $\delta \ne \delta_i \in \Delta$;
(ii) if $\delta_i < \delta < \delta'$ then $\delta' \prec_i \delta$;
(iii) if $\delta_i > \delta > \delta'$ then $\delta' \prec_i \delta$;
and (iv) $\prec_i$ is the smallest relation satisfying (i)-(iii).

Hence, a voter approves a proposals if and only if it is  in between the status quo and her ideal point, and 
the voter profile $V^{\delta,\delta'}$ on the amendment decision $d^\delta(\delta')$ for $\delta \neq \delta'$ is defined accordingly by $V^{\delta,\delta'}_i = 1$ iff $\delta_i \leq \delta' < \delta$ or $\delta_i \geq \delta' > \delta$.  

We can now specify formally Posterior Consistency:

\begin{enumerate}[itemsep=1ex]
\setcounter{enumi}{13}

\item \label{faxiom:posterior}
\textbf{Posterior Consistency}: Given an ideal points profile $\delta_i$, $i \in [n]$, and $\delta, \delta' \in \Delta$,
$d^\delta(V^{\delta,\delta'}) = 1$ only if $d^{\delta'}(V^{\delta,\delta'}) = 1$
\end{enumerate}

According to Axiom~\ref{faxiom:condorcet}, 
the amendment process is not a simple approval vote on a proposed amendment, but a selection of an most-preferred decision rule based on the reported ideal points of the agents.  Hence, the decision on an amendment is of a different type,
$\Delta \times \Delta^n \xrightarrow{} \Delta$, which takes the $\delta$ in force and the ideal points profile $\delta_i$, $i \in [n]$, as input, and produces the amended $\delta'$ as output.

Recall that a proposal $p$ is \emph{preferred} over  (aka \emph{dominates}) another proposal $p'$ if 
$|\{i \in [n] : p' \prec_i p\}|> \frac{n}{2}$. A proposal $p$ is \emph{most-preferred} (aka \emph{undominated}) if no other proposal $p'$ is preferred over $p$. With this we can specify Condorcet Consistency:

\begin{enumerate}[itemsep=1ex]
\setcounter{enumi}{14}

\item \label{faxiom:condorcet}
\textbf{Condorcet Consistency}: Given an ideal points profile $\delta_i$, $i \in [n]$,  choose a most-preferred\footnote{The standard definition of Condorcet Consistency is the selection of the unique Condorcet winner (namely undominated alternative) when it exists. Our definition is a bit more general.} 
$\delta' \in \Delta$  for which $d^\delta(V^{\delta,\delta'}) = 1$.
\end{enumerate}
In other words, given the current $\delta$-supermajority approval rule, Condorcet Consistency requires the choice of an amendment $\delta'$ that maximizes $\size{V^{\delta,\delta'}}$.

We can now state our main theorem.

\begin{theorem}[Condorcet Amendment]\label{theorem:optimal}
Given ideal points profile $\delta_i$ for $i \in [n]$, Axioms~\ref{faxiom:decisiveness}-\ref{faxiom:condorcet} and our assumptions on agent preferences imply that $d^\delta$ should be amended to $d^{\delta'}$ for $\delta' \neq \delta$ if either:
\begin{enumerate}[itemsep=1ex]
    \item  $\delta' = \argmax\limits_{\delta < x < 1} |\{i \in [n] : \delta_i \geq x\}| > x n$ exists, or\\

    \item $\delta' = \argmin\limits_{1/2 \le x < \delta} |\{i \in [n] : \delta_i \leq x\}| > \delta n$ exists
\end{enumerate}
Otherwise, no proposal to amend the decision rule is accepted. 
\end{theorem}

\begin{remark}
Note that the two definitions of $\delta'$ could be made more similar textually by replacing the right-hand side of the equations with $> max(x,\delta) \cdot n$.
\end{remark}

\begin{proof}
First, recall that, in all cases, the result is a $\delta$-supermajority rule, which is consistent with Axioms~\ref{faxiom:decisiveness}-\ref{faxiom:bootstrap}.  Next we argue that the $\delta'$ computed by the rules above is uniquely consistent with Axioms~\ref{faxiom:decisiveness}-\ref{faxiom:condorcet}:

\begin{enumerate}[itemsep=1ex]
    \item  If such a $\delta'>\delta$ exists, then $d^\delta$ would accept a proposal to amend $\delta$ to $\delta'$ since $\delta'$ is preferable to $\delta$ by a $\delta'$-supermajority, which is more than a $\delta$-supermajority. Furthermore, it is Posterior Consistent as the proposal to amend $\delta$ to $\delta'$ will be approved by $d^{\delta'}$ for the same reason.  
    Now we argue that $\delta'$ uniquely satisfies the axioms. A $\delta''> \delta'$ cannot be chosen without violating Posterior Consistency, since there is no $\delta''$-supermajority approval for that amendment by the maximality of $\delta'$. 
    A $\delta''<\delta'$ cannot be chosen without violating Condorcet consistency, since $\delta'$ is preferred over $\delta''$ by a $\delta'$-supermajority and hence by a majority.

    \item If such a $\delta'<\delta$ exists, note that accepting $d^{\delta'}$ as the new decision rule satisfies $d^\delta$ by the definition of $\delta'$, namely the use of $\delta n$ on the right hand side.
    Furthermore, since $\delta'$ is preferable to $\delta$ by a $\delta$-supermajority, which is greater than a $\delta'$-supermajority, it is also Posterior Consistent. 
    We now argue that $\delta$ uniquely satisfies the axioms. A $\delta''< \delta'$ cannot be accepted by $d^\delta$,  since there is no $\delta$-supermajority approval for amending $\delta$ to $\delta''$, by the minimality of $\delta'$. 
    And a $\delta''>\delta'$ cannot be chosen without violating Condorcet consistency, since $\delta'$ is preferred over $\delta''$ by a $\delta'$-supermajority and hence by a majority.

    \item If there is no $\delta$-supermajority to increase or decrease $\delta$ then of course $\delta$ should not change. No amendment proposal could be accepted according to $\delta$ itself, so remaining at $\delta$ in the face of any proposal is Posterior Consistent and trivially Condorcet-consistent.
\end{enumerate}
This finishes the proof of Theorem~\ref{theorem:optimal}.
\qed\end{proof}

We have shown that a group of $n$ agents that agree on a simple and self-evident axioms and a minimal set of assumption, can obtain from these a definite decision rule that also determines its own amendment process, via formal logical deduction.

\section{Alternative Approaches: Conservative Amendment}\label{section:alternative}

Naturally, there are many possible alternatives to the axiomatic basis proposed herein, as well as to the assumptions made.
Here we explore an alternative to Axiom \ref{axiom:condorcet}.

\begin{enumerate}
\setcounter{enumi}{6}
\item \label{axiom:conservatism}
\textbf{a. Conservatism}: Among the decision rules that satisfy all axioms above, choose the one closest to the current decision rule.
\end{enumerate}

Given our assumptions on agent preferences and Axioms~\ref{axiom:decisiveness}-\ref{axiom:bootstrap}, we also claim that there is a unique amendment process that is consistent with Axioms~\ref{axiom:decisiveness}-\ref{axiom:conservatism}(a). 

In practice, Condorcet Amendment happens to select the furthest $\delta$-supermajority rule from the current rule that satisfies Axioms~\ref{axiom:decisiveness}-\ref{axiom:posterior}. By contrast, Conservative Amendment requires that the rule selected be as close to the current rule as possible.

\begin{informalclaim}[Conservative Amendment]\label{claim:conservative}
Axioms~\ref{axiom:decisiveness}-\ref{axiom:conservatism}(a) together with our assumptions on agent preferences imply that the process to amend the current $\delta$-supermajority approval decision rule must be:
\begin{enumerate}[itemsep=1ex]
\item Increase $\delta$ to $\delta'>\delta$, if and only if there is a $\delta'$-supermajority that prefers $\delta'$ over $\delta$ and $\delta'$ is the \textbf{minimal} supermajority with this property; 
\item Decrease $\delta$ to $\delta'<\delta$, if and only if there is a $\delta$-supermajority that prefers $\delta'$ over $\delta$ and $\delta'$ is the \textbf{maximal} supermajority with this property; and 
\item  Retain the present $\delta$-supermajority rule otherwise.
\end{enumerate}
\end{informalclaim}

While the Condorcet and Conservative Amendment rules seem rather different, they are related in the sense that the second leads to the same result of the first if applied iteratively, in an exhaustive way -- in case the rule in force is simple majority.

\begin{informalclaim}[Iterate from Simple Majority]\label{claim:equivalence}
For any given ideal points profile, if the rule in force is $\delta = \frac{1}{2}$ then the iterative application of the Conservative Amendment rule until no further amendments occur halts at the decision rule characterized by Informal Claim \ref{claim:condorcet}.
\end{informalclaim}

To recap the informal presentation, if agents found their constitution according to Axioms~\ref{axiom:decisiveness}-\ref{axiom:bootstrap}, their initial decision rule will be simple majority. If they amend their initial rule by applying iteratively either of two amendment rules that satisfy posterior consistency, with a given set of ideal points,  the result would be the $\delta$ characterized by Informal Claim \ref{claim:condorcet}.

We now proceed with the formal analysis, considering Conservatism instead of Condorcet Consistency. 
First, we offer a formal statement of Conservatism:

\begin{enumerate}
\setcounter{enumi}{14}
\item \label{faxiom:conservatism} 
\textbf{a. Conservatism}: Given an ideal points profile $\delta_i$, $i \in [n]$,  choose a $\delta'\ne \delta \in \Delta$ closest to $\delta$ for which $d^\delta(V^{\delta,\delta'}) = 1$.
\end{enumerate}

The resulting Theorem \ref{theorem:conservative} looks very similar to Theorem~\ref{theorem:optimal}, but the two cases swap the minimization and maximization over values of $\delta$. The proof is similar as well.

\begin{theorem}[Conservative Amendment]\label{theorem:conservative}
Given ideal points $\delta_i$ for $i \in [n]$, Axioms~\ref{faxiom:decisiveness}-\ref{faxiom:conservatism}(a) imply that $d^\delta$ should be amended to $d^{\delta'}$ for $\delta' \neq \delta$ if either:
\begin{enumerate}[itemsep=1ex]
    \item  $\delta' = \argmin\limits_{\delta < x < 1} |\{i \in [n] : \delta_i \geq x\}| > x n$ exists, or\\

    \item $\delta' = \argmax\limits_{1/2 \le x < \delta} |\{i \in [n] : \delta_i \leq x\}| > \delta n$ exists
\end{enumerate}
Otherwise, no proposal to amend the decision rule is accepted. 
\end{theorem}

\begin{proof}
First, recall that in all cases the result is a $\delta$-supermajority rule, which is consistent with Axioms~\ref{faxiom:decisiveness}-\ref{faxiom:bootstrap}.  Next we argue that the $\delta'$ computed by the rules above is uniquely consistent with Axioms~\ref{faxiom:decisiveness}-\ref{faxiom:conservatism}(a):

\begin{enumerate}[itemsep=1ex]
    \item  If such a $\delta'>\delta$ exists, then $d^\delta$ would accept a proposal to amend $\delta$ to $\delta'$ since $\delta'$ is preferable to $\delta$ by a $\delta'$-supermajority, which is more than a $\delta$-supermajority. Furthermore, it is Posterior Consistent as the proposal to amend $\delta$ to $\delta'$ will be approved by $d^{\delta'}$ for the same reason.  
    We now show that $\delta'$ uniquely satisfies the axioms. Any $\delta'' > \delta'$ cannot be chosen without violating Conservatism because $\delta'$ would be accepted. No $\delta'' < \delta'$ could be accepted, because this would imply that $\delta''$ is preferred by a $\delta''$-supermajority according to Posterior Consistency, which violates the definition of $\delta'$.

    \item If such a $\delta' < \delta$ exists, note that accepting $d^{\delta'}$ as the new decision rule satisfies $d^\delta$ from the definition of $\delta'$, namely the use of $\delta n$ on the right hand side.
    Furthermore, since $\delta'$ is preferable to $\delta$ by a $\delta$-supermajority, which is greater than a $\delta'$-supermajority, it is also Posterior Consistent. 
    We now argue that $\delta'$ uniquely satisfies the axioms. 
    A $\delta''>\delta'$ cannot be chosen because according to Posterior Consistency it must be preferred by a $\delta''$-supermajority, which violates the definition of $\delta'$. 
    A $\delta''< \delta'$ cannot be accepted without violating Conservatism because $\delta'$ would be accepted. 
    
    \item If there is no $\delta$-supermajority to increase or decrease $\delta$ then of course $\delta$ should not change. No amendment proposal could be accepted according to $\delta$ itself, so remaining at $\delta$ is Posterior Consistent and trivially Conservative.
\end{enumerate}
This finishes the proof of Theorem~\ref{theorem:conservative}.
\qed\end{proof}


\begin{proposition}[Increasing Condorcet Amendment is Idempotent]\label{proposition:idempotence}
Given the initial decision rule $\delta = \frac{1}{2}$ and any ideal points profile, the application Condorcet Consistent Amendment is idempotent.
\end{proposition}
\begin{proof}
When the current rule is $\delta = \frac{1}{2}$, any change to the decision rule must increase $\delta$. From Theorem~\ref{theorem:optimal}, under Condorcet Amendment, $\delta'$ will be the largest such that the number of agents with ideal points at least $\delta'$ is strictly greater than $\delta' n$. Let us refer to this decision rule as $\hat{\delta} \in \Delta$. Since a $\delta'$-supermajority prefer $\hat{\delta}$ over $\delta$, we know the majority cannot prefer any $\delta'' < \hat{\delta}$ over $\hat{\delta}$, so once $\hat{\delta}$ takes over as the current decision rule, no proposal to decrease it will be accepted. For an amendment $\delta'' > \hat{\delta}$ to occur, there would have to be a $\delta''$-supermajority who prefer $\delta''$ over $\hat{\delta}$, but this violates the definition of $\hat{\delta}$. Thus, Condorcet Amendment is idempotent
and chooses $\hat{\delta}$.
\end{proof}

\begin{proposition}[Iterate from Simple Majority]\label{proposition:equivalence}
Given the initial decision rule $\delta = \frac{1}{2}$ and any ideal points profile, the result of iterative application to completion of Conservative Amendment is the same as a single application of Condorcet Amendment.
\end{proposition}

\begin{proof}
Suppose that the iterative application of Conservative Amendment were to select some $\delta' \neq \hat{\delta}$.
First, suppose that $\delta' < \hat{\delta}$.
A $\hat{\delta}$-supermajority prefers $\hat{\delta}$ to~$\delta'$, which is larger than a $\delta$-supermajority, so Conservative Amendment, given $\delta'$, would approve the proposal $\hat{\delta}$. This contradicts the fact that the iterative application of Conservative Amendment reached completion.
Second, suppose $\delta' > \hat{\delta}$. From Theorem~\ref{theorem:conservative}, this implies that more than $\delta' n$ agents have ideal points of at least $\delta'$, because it is accepted when proposed against a smaller $\delta$. This contradicts the definition of $\hat{\delta}$.
\qed\end{proof}

Here, we have offered an alternative axiomatic basis, replacing Condorcet Consistency (Axioms \ref{axiom:condorcet}, \ref{faxiom:condorcet}) with Conservatism (Axioms \ref{axiom:conservatism}(a) and \ref{faxiom:conservatism}(a)), discussed its consequences, and compared it to the initial axiomatic basis proposed.

\section{Discussion}\label{section:discussion}

We considered a set of agents forming an egalitarian, self-governed community needing to establish by what process they will make decisions.

We started by assuming that the agents agree on a small set of intuitive axioms, and showed that from these axioms alone arises a simple constitution -- an initial decision rule for making decisions on whether to accept or reject proposals, which can be applied to itself if the agents wish to change the rule.
One of the axioms, Minimality, was unique among the axioms because it compared possible rules and prioritized one over the rest rather than restricting the set of possible rules. There are possible replacements to consider, including Unanimity.

We have argued that a rule that amends itself requires additional considerations, and  offered additional axioms for that case. We have shown that basic assumptions on the structure of agent preferences over possible rules and these axioms result in a unique amendment process.  We have considered two alternatives for the final axiom - Condorcet Consistency and Conservatism.
While Condorcet Consistency results in a different amendment process from Conservatism, we have shown that repeated application of the Conservatism process, starting from the initial simple majority rule, results in the same rule produced by the Condorcet Consistent amendment process.

One natural next step following our work is to consider different structures of agent preferences over rules (e.g., metric preferences) with the same axioms. A second is to determine what alternative sets of axioms might lead to different sets of rules that can also be used to amend themselves.

Lastly, a key assumption we have made here is that the set of agents is fixed. However, any self-governed community must determine who its members are when founding (i.e. whose votes count), and must decide for itself how to add and remove members in the future~\cite{alcantud2018collective,alcantud2020independent,cho2020group,danezis2009sybilinfer,dimitrov2007procedural,fioravanti2020asking,kasher1997question,miller2008group,poupko2019sybil,poupko2021building,shahaf2019genuine,sung2005axiomatic,wei2012sybildefender,yu2008sybillimit,yu2008sybilguard}. 

\section*{Acknowledgements}
Ehud Shapiro is the Incumbent of The Harry Weinrebe Professorial Chair of Computer
Science and Biology. We thank the generous support of the Braginsky Center for the
Interface between Science and the Humanities. Nimrod Talmon was supported by the
Israel Science Foundation (ISF; Grant No. 630/19).
Ben Abramowitz was supported in part by NSF award CCF-1527497.

\bibliographystyle{plain}
\bibliography{bib}

\end{document}